\documentclass{mnras}

\usepackage{natbib}
\usepackage{graphicx}
\usepackage{epstopdf}
\usepackage[T1]{fontenc}
\usepackage{aecompl}
\usepackage{mnras-bst}
\usepackage{hyperref}
\bibpunct{(}{)}{;}{a}{}{,}

\usepackage{amsmath}

\newcommand\ksi{\xi}
\newcommand\0{\phantom{0}}

\title[Polarization of $\ksi$~Boo~A]{The Rotationally Modulated Polarization of $\ksi$~Boo~A}
\author[D.V.~Cotton et al.]{Daniel V. Cotton$^{1,2}$\thanks{E-mail: d.cotton@unsw.edu.au}, 
Dag Evensberget$^{3}$, Stephen C. Marsden$^{3}$, Jeremy Bailey$^{1,2}$, 
\newauthor Jinglin Zhao$^{1}$, Lucyna Kedziora-Chudczer$^{1,2}$, Bradley D. Carter$^{3}$, Kimberly Bott$^{4,5}$,
\newauthor Aline A. Vidotto$^{6}$, Pascal Petit$^{7,8}$, Julien Morin$^{9}$ and Sandra V. Jeffers$^{10}$.\\
\\
$^{1}$School of Physics, UNSW Sydney, NSW 2052, Australia.\\
$^{2}$Australian Centre for Astrobiology, UNSW Sydney, NSW 2052, Australia. \\
$^{3}$University of Southern Queensland, Centre for Astrophysics, Springfield, Qld. 4300/Toowoomba, Qld. 4350, Australia. \\
$^{4}$University of Washington Astronomy Department, Box 351580, UW Seattle, WA 98195, USA. \\
$^{5}$NExSS Virtual Planetary Laboratory, Box 351580, UW Seattle, WA 98195, USA.\\
$^{6}$School of Physics, Trinity College Dublin, College Green, Dublin 2, Ireland.\\
$^{7}$Universit\'e de Toulouse, UPS-OMP, IRAP, Toulouse, France.\\
$^{8}$CNRS, Institut de Recherche en Astrophysique et Planetologie, 14, avenue Edouard Belin, F-31400 Toulouse, France.\\ 
$^{9}$LUPM-UMR 5299, CNRS \& Universit\'e Montpellier, place Eug\`ene Bataillon, 34095 Montpellier Cedex 05, France.\\
$^{10}$Institute for Astrophysics, University of Goettingen, Friedrich Hund Platz 1, 37077, Goettingen, Germany.}
\begin{document}

\date{Accepted . Received ; in original form }

\pagerange{\pageref{firstpage}--\pageref{lastpage}} \pubyear{2018}

\maketitle

\label{firstpage}

\begin{abstract} We have observed the active star $\ksi$~Boo~A (HD~131156A) with high precision broadband linear polarimetry contemporaneously with circular spectropolarimetry. We find both signals are modulated by the 6.43~day rotation period of $\ksi$~Boo~A. The signals from the two techniques are 0.25 out of phase, consistent with the broadband linear polarization resulting from differential saturation of spectral lines in the global transverse magnetic field. The mean magnitude of the linear polarization signal is $\sim$4~ppm/G but its structure is complex and the amplitude of the variations suppressed relative to the longitudinal magnetic field. The result has important implications for current attempts to detect polarized light from hot Jupiters orbiting active stars in the combined light of the star and planet. In such work stellar activity will manifest as noise, both on the time scale of stellar rotation, and on longer time scales -- where changes in activity level will manifest as a baseline shift between observing runs. \end{abstract}


\section{Introduction}

The primary mode of characterising the magnetic field in a star is through circular polarimetry. In highly magnetic stars linear polarization may be used to complement measurements of circular polarization, and constrain magnetic field geometry (\citealp{wade00} and references therein). However, modern stellar polarimeters are much more sensitive to circular polarization than to the inherently weaker signal from linear polarization. Only in the last decade has linear polarization been definitively detected in (bright) weakly magnetic stars \citep{kochukhov10,rosen15}. The difference arises as a result of the polarimetric mechanism. In a magnetic field circular polarization is produced by the longitudinal Zeeman effect splitting spectral lines into two oppositely polarized (left and right handed) lines. Linear polarization is produced by the transverse Zeeman effect splitting lines in three, where the outside lines are polarized in one orientation and the centre line -- having double the intensity -- is polarized in the other \citep{stenflo13}. When the magnetic field is weak, the lines are not completely split, but instead the two components are to be found predominantly in the line wings and line core respectively. Spectropolarimetry -- where the line profiles are fit to determine polarization and hence magnetic field strength -- can be used to measure both types of polarization, but the line profiles of circular polarization are much more easily detected \citep{wade00}.

In our recent paper we measured significant \textit{broadband} linear polarization in a number of active late-type dwarf stars -- mostly BY~Dra variables and stars with emission line spectral types \citep{cotton17a}. In stars with very strong magnetic fields a net linear polarization will be measured in a line when the line core is saturated -- called magnetic intensification \citep{babcock49}. Similarly, `differential saturation' describes the situation where many lines overlap and merge with each other (line blanketing) to produce a net broadband linear polarization \citep{bagnulo95}. The broadband linear polarization magnitude measured in the active dwarfs was correlated with the maximum \textit{global} longitudinal magnetic field (|$B_\ell$|$_{max}$) from spectropolarimetric (circular polarimetry) measurements. Consequently it is presumed that the broadband linear polarization measured in these active dwarfs is produced through differential saturation that is also induced by the global magnetic field. If so, the field geometry will be important, a uniform dipolar field aligned with the stellar rotation axis might produce a constant polarization, more complicated structures will result in a time varying signal. However, linear polarization may also be generated in active stars through other mechanisms with more complicated phase behaviour. Strong localised fields might be produced in starspots \citep{huovelin91,saar93}. Or starspots might produce polarization by breaking symmetry, not in the spectral lines, but on the disk of the star instead \citep{yakobchuk18}. In red super/giant stars, stellar hotspots have also been found to produce linear polarization \citep{schwarz86,auriere16}.

Determining the polarimetric mechanism of the linear polarization in active dwarfs is important, not just for the information complementary to spectropolarimetry (e.g. \citealp{wade96,rosen15}), but also because it is a potential source of noise in studying other polarimetric phenomena. In particular, a number of groups have been searching for the polarized light that is scattered from the atmosphere of a close hot Jupiter planet, in the combined light of the star and the planet \citep{berdyugina11,lucas09,wiktorowicz15b,bott18b}. If identified, such a signal can reveal details of the planet's atmosphere: its albedo and cloud properties. However, stellar activity is likely to significantly complicate such searches as the expected signal due to an orbiting, unresolved exoplanet is smaller than that seen in active dwarfs \citep{seager00,bailey18}, and many of the best candidate systems for detecting a planetary signal (those with very short period planets orbiting bright stars) have late type dwarf star hosts that are active or potentially active. If activity effects are to be avoided, or removed, it is important they be understood. 

As the most polarized star identified in \citet{cotton17a} $\ksi$~Boo~A (HD~131156A) is the most obvious candidate to look for and characterise any variability. It is a BY~Dra variable star \citep{samus03} with a G7Ve spectral type \citep{levato78} lying 6.7~pc from the Sun \citep{vanleeuwen07}. It has a close companion which was 5.41$\arcsec$ away (at the time of our observations), $\ksi$~Boo~B (HD~131156B), which is also an active star, of spectral type K5Ve \citep{levato78}. $\ksi$~Boo~A has a short, 6.43~d, rotational period \citep{toner88}. In an early study \citet{huovelin88} concluded that $\ksi$~Boo~A varies in linear polarization around its rotational cycle based on data greater than 2-$\sigma$ from the mean. 

\citet{petit05} found $\ksi$~Boo~A has a magnetic field made up of two main components: a 40~G dipole inclined at 35$\degr$ to the rotation axis, and a large-scale 120~G toroidal field. Twenty years of data presented by \citet{lockwood07} shows that its activity can vary on long timescales in a seemingly irregular way. Similarly \citet{morgenthaler12} present field maps corresponding to the years 2007 to 2011 that show varying behaviour. On a shorter time scale, in 101 measurements, the BCool team \citep{marsden14} determined |$B_\ell$|$_{max}$ as 18.4~$\pm$~0.3~G and |$B_\ell$|$_{min}$ as 0.5~$\pm$~1.0~G. This is quite a strong field compared to other Solar type stars, but is still a weak field compared to the fields found within star/sunspots or those in the hotter stars where linear polarimetry has traditionally been employed \citep{wade96}.

\section{Observations}
\label{sec:obs}

\subsection{Linear Polarimetry}

Broadband linear polarization measurements were made with the HIgh Precision Polarimetric Instrument (HIPPI, \citealp{bailey15}) on the 3.9-m Anglo-Australian Telescope, at Siding Spring Observatory in Australia. The instrument was mounted at the F/8 Cassegrain focus, giving an aperture size of 6.7$\arcsec$ -- just small enough to isolate $\ksi$~Boo~A from $\ksi$~Boo~B at the time of our observations (it is difficult to measure the seeing with HIPPI on the telescope accurately, but the seeing was decent -- generally around 2$\arcsec$ or better -- for our observations). HIPPI has a night-to-night precision of 4.3~ppm on bright stars, which it achieves using a (Boulder Non-linear Systems) ferro-electric liquid crystal modulator operating at 500~Hz, and two additional slower stages of chopping \citep{bailey15}. We configured HIPPI to use Hamamatsu H10720-210 ultra bi-alkali photocathode photomultiplier tubes as detectors, and no photometric filter (Clear) -- giving flux between 350~nm and 730~nm. This is the usual configuration of the instrument for observation of exoplanet systems, (e.g. the (inactive) WASP-18 system, \citealp{bott18b}). For the G7 spectral type of $\ksi$~Boo~A the effective wavelength is 486.1~nm and the modulation efficiency 0.840.

We observed $\ksi$~Boo~A during two observing runs: June/July 2017 and August 2017. The telescope polarization (TP) is stable over such a time frame \citep{cotton16a,marshall16,cotton17a,cotton17b}, and was determined by taking the mean of all low polarization standard star observations. These are shown as normalised Stokes parameters in table \ref{tab:tp}, where $q=Q/I$ and $u=U/I$; the total linear polarization can be calculated as $p=\sqrt{q^2+u^2}$. 

The position angle (PA) was calibrated by reference to standard stars: HD~147084 (twice), HD~154445 and HD~160529 in June/July; and HD~147084, HD~154445 and HD~187929 in August. The $\sim$1$\degr$ error in the PA determination is dominated by the uncertainties in the PAs of the standards (see \citealp{cotton17b} for standard details).


\begin{table}
\caption{TP determination from low polarization standard observations. Exposure times are 320~s for Sirius and 640~s otherwise.}
\tabcolsep 3 pt
\centering
\begin{tabular}{lcrr}
\hline
Star        &   UTC                      &   $q$ (ppm)  &   $u$ (ppm)  \\
\hline
$\beta$ Hyi	&   2017-06-22 19:21:20	    &   $-$25.9~$\pm$~\03.4   &   $-$0.9~$\pm$~\03.3 \\
        	&   2017-06-29 19:38:38	    &   $-$22.1~$\pm$~\05.0	    &    1.1~$\pm$~\04.7 \\
        	&   2017-06-30 19:50:35	    &   $-$15.8~$\pm$~\03.9	    &   11.7~$\pm$~\03.8 \\
            &   2017-08-11 16:26:58	    &   $-$21.5~$\pm$~\04.6	    &   21.4~$\pm$~\04.4 \\
            &   2017-08-11 17:48:42     &	$-$19.0~$\pm$~\03.9	  &   $-$3.7~$\pm$~\04.0 \\
            &   2017-08-17 19:24:21     &	$-$31.5~$\pm$~\04.7	    &    2.4~$\pm$~\05.2 \\
Sirius      &	2017-08-10 19:49:29	    &   $-$18.6~$\pm$~\03.6     &	 6.2~$\pm$~\03.9 \\
            &   2017-08-15 19:23:35     &	$-$25.5~$\pm$~\08.3   &  $-$15.7~$\pm$~\07.9 \\
            &   2017-08-16 19:27:41     &	$-$19.5~$\pm$~\02.7   &	    20.6~$\pm$~\03.6 \\
            &   2017-08-18 19:07:11     &	$-$10.1~$\pm$~12.2    &	    12.8~$\pm$~13.9 \\
            &   2017-08-19 19:33:56     &	    3.0~$\pm$~\02.3   &	  $-$9.0~$\pm$~\02.5 \\
$\beta$ Leo &   2017-06-22 08:17:33	    &    $-$3.1~$\pm$~\02.4   &	  $-$6.0~$\pm$~\02.4 \\
            &   2017-06-26 08:17:37	    &    $-$8.0~$\pm$~\02.4   &  $-$11.9~$\pm$~\02.2 \\
            &   2017-07-05 08:15:41     &	$-$10.1~$\pm$~\03.4   &	  $-$7.8~$\pm$~\02.9 \\
$\beta$ Vir &   2017-06-23 08:14:32 	&    $-$2.7~$\pm$~\05.1   &	  $-$7.6~$\pm$~\04.9 \\
            &   2017-06-24 08:20:36 	&    $-$8.8~$\pm$~\04.6	&     $-$5.7~$\pm$~\04.7 \\
\hline
Adopted TP  &                           &   $-$15.0~$\pm$~\00.3   &    0.5~$\pm$~\00.3 \\
\hline
\end{tabular}
\label{tab:tp}
\end{table}

Table \ref{tab:lp} gives the $\ksi$~Boo~A observations after PA calibration, subtraction of the TP and efficiency correction.  

\begin{table}
\caption{HIPPI observations of $\ksi$~Boo~A. Exposure time on 2017-08-14 was 1000~s, but 800~s otherwise. $\hat{p_{\star}}$ is the intrinsic linear polarization (see section \ref{sec:analysis}) debiased as $\hat{p}=\sqrt{p^2-\Delta p^2}$.}
\tabcolsep 4 pt
\centering
\begin{tabular}{crrr}
\hline
UTC                  &  $q$ (ppm) & $u$ (ppm) &  $\hat{p_{\star}}$ (ppm)\\
\hline
2017-06-22 11:34:57	&	54.8~$\pm$~\06.5	&	$-$20.9~$\pm$~\06.7   &   56.3~$\pm$~\07.3   \\
2017-06-24 11:14:51	&   23.3~$\pm$~\09.6	&	$-$28.4~$\pm$~11.5    &   61.9~$\pm$~10.4   \\
2017-06-25 10:37:15	&   41.0~$\pm$~\07.0	&	   15.5~$\pm$~\06.7   &   74.2~$\pm$~\08.0   \\
2017-06-26 09:34:48	&   44.2~$\pm$~\07.3	&	   10.3~$\pm$~\07.1   &   45.2~$\pm$~\09.4   \\
2017-06-29 11:45:04	&   42.0~$\pm$~\08.6	&	$-$23.3~$\pm$~\09.1   &   66.7~$\pm$~\08.3   \\
2017-06-30 11:07:24	&   35.9~$\pm$~\07.2	&	$-$27.1~$\pm$~\07.4   &   37.8~$\pm$~\08.2   \\
2017-07-01 11:06:51	&   29.0~$\pm$~\06.6	&	    1.4~$\pm$~\06.7   &   42.4~$\pm$~\08.0   \\
2017-07-02 11:30:35	&   36.3~$\pm$~\06.6	&	 $-$5.7~$\pm$~\06.7   &   33.2~$\pm$~11.0   \\
2017-07-05 11:44:28	&   68.1~$\pm$~\07.3	&	$-$34.9~$\pm$~\07.4   &   18.5~$\pm$~\08.2   \\
2017-08-10 09:48:23	&   39.5~$\pm$~\07.0	&	    5.7~$\pm$~\06.9   &   26.4~$\pm$~\07.4   \\
2017-08-12 09:48:29	&   39.9~$\pm$~\07.6	&	   11.2~$\pm$~\07.3   &   41.9~$\pm$~\07.6   \\
2017-08-14 09:47:14	&   34.1~$\pm$~\07.6	&	$-$22.0~$\pm$~\07.6   &   32.1~$\pm$~\07.6   \\
2017-08-15 09:45:45	&   18.8~$\pm$~\07.7	&	$-$11.7~$\pm$~\07.5   &   34.2~$\pm$~\07.4   \\
2017-08-16 09:49:53	&   34.7~$\pm$~\06.8	&	 $-$1.1~$\pm$~\06.9   &   43.2~$\pm$~\07.9   \\
2017-08-19 09:50:30	&   55.7~$\pm$~10.1	    &	$-$32.9~$\pm$~\09.6   &   37.6~$\pm$~\07.7   \\
2017-08-20 08:40:44	&   65.8~$\pm$~\07.7	&	$-$21.0~$\pm$~\07.7   &   39.2~$\pm$~\08.1   \\
\hline
\end{tabular}
\label{tab:lp}
\end{table}

\subsection{Circular Spectropolarimetry}
\label{sec:spectro}

Spectropolarimetric observations were made with the NARVAL \'{e}chelle spectrometer operating at the T\'{e}lescope Bernard Lyot (Observatoire du Pic du Midi, France). NARVAL \citep{donati06} is a bench mounted spectrograph connected by optical fibre to a Cassegrain mounted polarimetric module. The polarimetric module comprises a series of Fresnel rhombs and a Wollaston prism. The configurations permits the simultaneous recording of Stokes $I$ and $V$ spectra. Each measurement comprises four exposures with different half-wave rhomb orientations to remove instrument effects \citep{semel93,donati97}. NARVAL covers the wavelength range 370 to 1100~nm and has \textit{R} $\sim$65000, corresponding to a pixel size of $\sim$1.8~km/s in velocity space.




\begin{table}
\centering 
\caption{NARVAL observations of $\ksi$~Boo~A and derived quantities. Lines is the number of spectral lines with sufficient signal-to-noise ratio to be included when generating the LSD profile. The average Land\'{e} factor $\left<g\right>$ and average central wavelength of lines used $\left<\lambda\right>$ when generating the LSD profile are used to calculate the mean longitudinal magnetic field $B_\ell$, see equation \eqref{eq:B_ell}.}
\tabcolsep 5 pt
\centering
\begin{tabular}{cccccc}
\hline
UTC &  Lines &  $\left<g\right>$ &  $\left<\lambda\right>$ (nm) &  $B_\ell$ (G) \\
\hline
2017-06-02 00:40:21 &  10769 &             1.215 &                        560.9 &      +~11.8~$\pm$~0.7 \\
2017-06-05 21:54:06 &  10773 &             1.215 &                        547.5 &     +~\08.3~$\pm$~3.7 \\
2017-06-06 21:24:20 &  10772 &             1.215 &                        544.0 &      +~11.4~$\pm$~0.5 \\
2017-06-07 22:08:40 &  10779 &             1.215 &                        546.7 &      +~15.7~$\pm$~0.5 \\
2017-06-09 22:23:55 &  10773 &             1.214 &                        550.7 &     +~\03.2~$\pm$~1.1 \\
2017-06-10 22:24:20 &  10774 &             1.214 &                        549.5 &     +~\05.8~$\pm$~0.8 \\
2017-06-11 23:24:59 &  10772 &             1.215 &                        551.5 &     +~\09.3~$\pm$~0.6 \\
2017-06-12 22:22:54 &  10779 &             1.215 &                        547.6 &     +~\09.9~$\pm$~0.6 \\
2017-06-13 23:15:23 &  10773 &             1.215 &                        548.3 &      +~16.2~$\pm$~0.7 \\
2017-06-14 22:21:13 &  10773 &             1.215 &                        549.6 &      +~13.5~$\pm$~0.5 \\
2017-06-16 22:42:40 &  10778 &             1.215 &                        546.1 &     +~\07.9~$\pm$~0.5 \\
2017-06-17 22:33:12 &  10771 &             1.215 &                        545.9 &     +~\08.6~$\pm$~0.5 \\
2017-06-18 22:34:51 &  10771 &             1.215 &                        547.5 &     +~\09.5~$\pm$~0.5 \\
2017-07-02 22:13:38 &  10775 &             1.215 &                        555.7 &      +~12.3~$\pm$~0.7 \\
2017-07-05 22:15:22 &  10765 &             1.215 &                        549.8 &     +~\03.2~$\pm$~0.6 \\
\hline
\end{tabular}
\label{tab:BCoolObservations}
\end{table}


$\ksi$~Boo~A was observed 15 times with NARVAL (table \ref{tab:BCoolObservations}). For each observation, the resulting Stokes $I$ and $V$ spectra were reduced using Least-Squares Deconvolution (LSD) with \textsc{libre-esprit} (\citealp{donati97} describes \textsc{esprit}). The LSD technique transforms the spectra around a set of known spectral lines from wavelength space to velocity space and co-adds them to form one Stokes $I$ and one $V$ line profile parametrised by Doppler velocity, each with a high S/N.

As in the BCool survey \citep{marsden14}, $\ksi$~Boo~A stellar parameters from \cite{valenti05,takeda07}: $T_\text{eff}=$~5570~$\pm$~31~K, $\log{g}=$~4.57~$\pm$~0.02~cm/s$^2$
and $\log{M/H}=$~-0.07~$\pm$~0.02, were used to determine the closest of the BCool atmospheric line masks, defined by
$T_\text{eff}=$~5500~K, $\log(g)=$~4.5~cm/s$^2$ and $\log(M/H)=$~0, then used to generate the Stokes $I$ and $V$ LSD profiles. The line mask originates from the Vienna Atomic Line Database \citep{kupka00}.

The mean longitudinal magnetic field is the line-of-sight magnetic field component integrated over the stellar disk.  Following \citet{carroll14}, an estimate of this quantity can be calculated directly from the LSD line profile: \begin{equation}B_\ell = -\frac{h \int v V(v) \,\mathrm{d}v}{\mu_\text{B} \left<\lambda\right> \left<g\right> \int I_c - I(v) \,\mathrm{d}v}, \label{eq:B_ell} \end{equation} where $h$ is Planck's constant and $\mu_\text{B}$ is the Bohr magneton. The parameters $\left<g\right>$ and $\left<\lambda\right>$ are the average Land\'{e} factor and wavelength taken over the spectral lines used in forming the LSD profile, and are calculated by \textsc{libre-esprit} for each observation (see table \ref{tab:BCoolObservations}). We note that $hc/\mu_\text{B}=$~0.0214~Tm where $c$ is the speed of light, permitting the recovery of the form of equation \eqref{eq:B_ell} (see \citealp{donati97} where it is their equation 5, or \citealp{mathys89}).

Due to the presence of noise, the calculation of $B_\ell$ is affected by the range of velocities, $v$, over which the integrals in equation \eqref{eq:B_ell} are taken. The velocity range was chosen to maximise the periodic signal when calculating a Lomb-Scargle periodogram. The lowest false alarm probability of 3.0\% was found letting $v$ range over 11 velocity bins centered on $v=$~1.8~km/s. This gives a velocity line width for the Stokes $I$ and $V$ signals of $-$7.2 to 10.8~km/s. The centre agrees well with the BCool astrometric radial velocity of 1.9~km/s \citep{marsden14}.

\section{Analysis}
\label{sec:analysis}

The error weighted mean of the 16 HIPPI observations of $\ksi$~Boo~A is 41.4~$\pm$~1.9~ppm in $q$ and $-$9.1~$\pm$~1.9~ppm in $u$. This is very similar to our previously reported observations in the SDSS $g^{\rm \prime}$ band from February 2016 of 45.8~$\pm$~5.2 and 3.0~$\pm$~5.2~ppm respectively \citep{cotton17a}. The effective wavelength of HIPPI's clear band (486.8~nm) is similar to that of the $g^{\rm \prime}$ band (472.4~nm), however the clear band is much broader. Consequently, the difference in these mean measurements is not sufficient to demonstrate variability. 

The interstellar polarization calculated for $\ksi$~Boo~A is 1.7~$\pm$~3.2~ppm in $q$ and $-$0.9~$\pm$~3.2~ppm in $u$ \citep{cotton17a}, based on its distance from the Sun of 6.7~pc and the PA of nearby intrinsically unpolarized stars. Thus it is very likely the vast majority of polarization measured is intrinsic to the star. Subtracting the interstellar polarization and calculating the (debiased) linear polarization ($\hat{p}$) from $q$ and $u$ gives 40.5~$\pm$~3.7~ppm (we neglect the small effect from interstellar polarization colour -- see \citealp{marshall16} or \citealp{cotton18b} for a discussion of the wavelength dependence of interstellar polarization for nearby stars). The error weighted mean of $B_\ell$ from the contemporaneous NARVAL data is 10.7~$\pm$~0.2~G, so if the intrinsic linear polarization is due to the magnetic field the mean contribution is $\sim$3.8~ppm/G.  



\subsection{Linear polarimetry statistics}

The average error in our linear polarization data -- the internal standard deviation of individual measurements, which scales with photon-shot noise -- is 7.6~ppm in $q$ and 7.7~ppm in $u$. The standard deviation (or scatter, $\sigma$) in repeat observations in both $q$ and $u$ is higher than this, which may indicate intrinsic variability. The variability scale, sometimes called the error variance, is calculated as: \begin{equation}s = \sqrt{\sigma^2 - \sum_i \delta_i^2},\end{equation} where $\delta_i$ values are known errors, here the average error ($\bar{e}$), and the night-to-night precision of HIPPI. Thus the error variance is 10.1~ppm in $q$ and 13.8~ppm in $u$. By comparison the standard deviation of the spectropolarimetric data is 4.0~G. If we scale this value using the same ratio implied by the mean (3.8~ppm/G) this gives the equivalent of 15.2~ppm.


Table \ref{tab:stat} presents the standard deviation, skewness and kurtosis which characterise the variability in $q$ and $u$. By comparison with the tables of \citet{brooks94} the kurtosis in $u$ is non-Gaussian with 95\%, but not 99\% confidence -- indicative of a centre-heavy distribution -- but otherwise the data is consistent with being Gaussian.

\begin{table}
\caption{Statistical analysis of linear polarization measurements. The error weighted means ($\bar{x}_{wt}$), means ($\bar{x}$), mean errors ($\bar{e}$), standard deviations ($\sigma$), and error variances ($s$) are in ppm.}
\tabcolsep 3 pt
\centering
\begin{tabular}{cccccccccc}
\hline
Stokes  &   $\bar{x}_{wt}$       &   $\bar{x}$    &   $\bar{e}$  &   $\sigma$   &   $s$   &   Skewness    &   Kurtosis    \\
\hline
$q$ &   41.4~$\pm$~1.9  &   \phantom{+}41.4   &   7.6         &   13.4        &   10.1        &   0.2272      &   2.6180      \\
$u$ & $-$9.1~$\pm$~1.9  &  $-$11.6   &   7.7         &   16.4        &   13.8        &   0.0475      &   1.6130      \\
\hline
\end{tabular}
\label{tab:stat}
\end{table}

\subsection{Rotational modulation and the magnetic field}

\begin{figure*}
\centering
\includegraphics[width=0.75\textwidth]{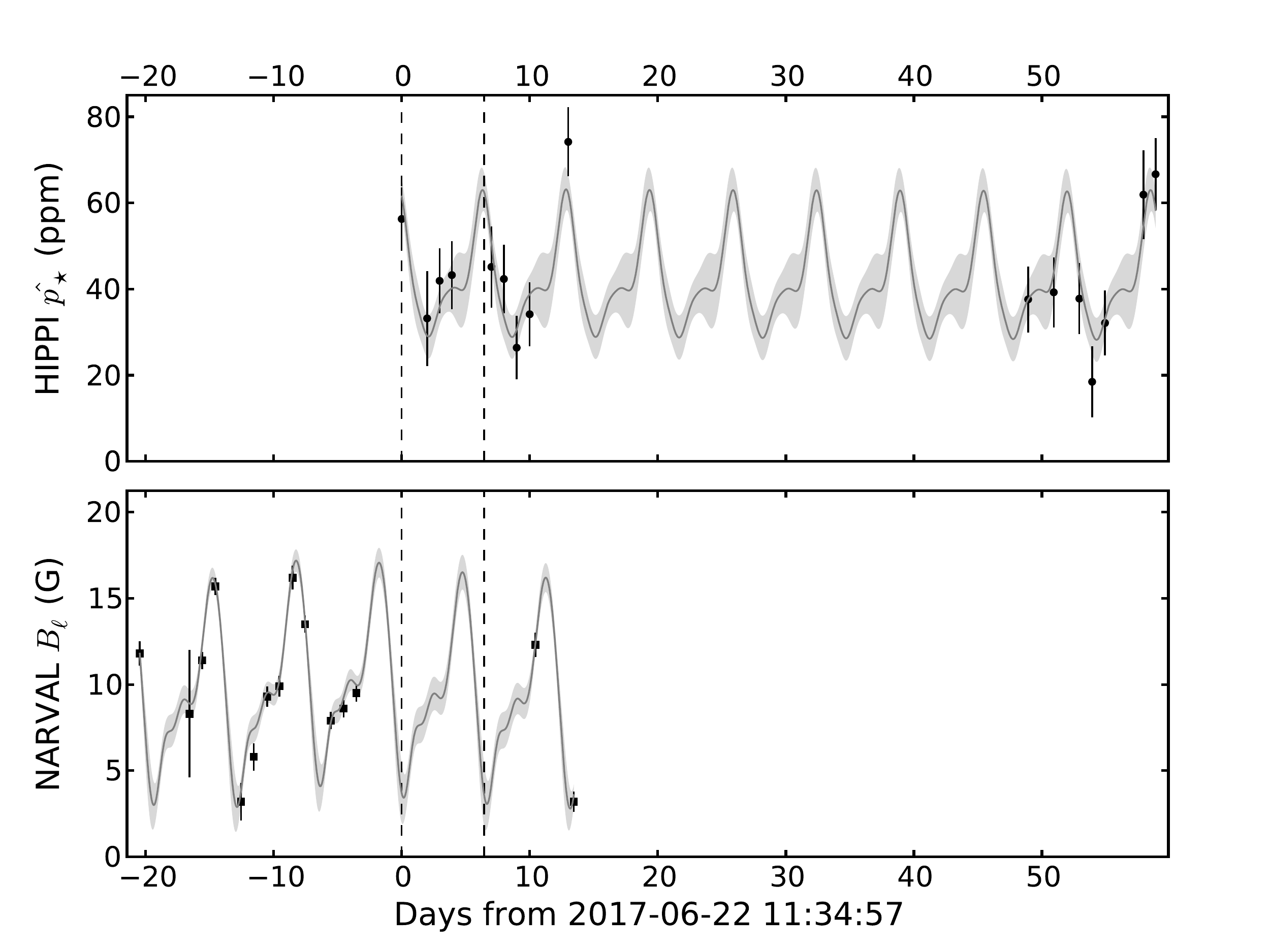}
\caption{The HIPPI $\hat{p_{\star}}$ (top) and NARVAL $B_\ell$ (bottom) data (black) and the Gaussian process maximum likelihood prediction (grey) (and 1-$\sigma$ errors -- light grey) used to determine periods. For comparison dashed lines show the chosen 6.43~d period.}
\label{fig:ml}
\end{figure*}

Several determinations of the rotational period of $\ksi$~Boo~A exist. Using longitudinal magnetic field measurements \citet{plachinda00} obtained 6.1455~$\pm$~0.0003~d. Activity indices based on Ca II H \& K lines have been used by \citet{noyes84}: 6.2~$\pm$~0.1, \citet{donahue96}: 6.31~d, and most recently \citet{hempelmann16}: 6.299~$\pm$~0.037. \citet{toner88} made a careful analysis of line symmetry and line ratios to get 6.43~$\pm$~0.01~d. The differences between these values may be related to their probing of different stellar layers, and the star's differential rotation as described by \citet{morgenthaler12}. We don't have sufficient data to determine the period so precisely ourselves, so use our own period analysis to inform a choice. We first constructed a Lomb-Scargle periodogram with \textsc{astropy}'s `LombScargle' package. Using the $\hat{p_{\star}}$ and $B_\ell$ data, we find respectively $\sim$6.5~d and 6.3~d. We also applied Gaussian process maximum likelihood estimation (figure \ref{fig:ml}), finding 6.52~$\substack{+0.05\\-0.08}$~d for $\hat{p_{\star}}$ and 6.50~$\pm$~0.06~d for $B_\ell$. For this we used a combination of the exponential squared kernel and the exp-sine-squared kernel to describe the quasi-periodicity of the magnetic field, as well as a rational-quadratic kernel that describes smooth signal changes at various time scales \citep{rasmussen06}. We employed the Python library \textsc{george} \citep{ambikasaran15} to implement Gaussian processes and estimated the uncertainty of the period using Markov chain Monte Carlo (MCMC) sampling with the \textsc{emcee} Python module \citep{foreman-mackey13}. Consequently, as the best match for our data, we prefer the 6.43~d period of \citet{toner88} -- as did \citet{petit05}.


\begin{figure*}
\centering
\includegraphics[width=0.75\textwidth]{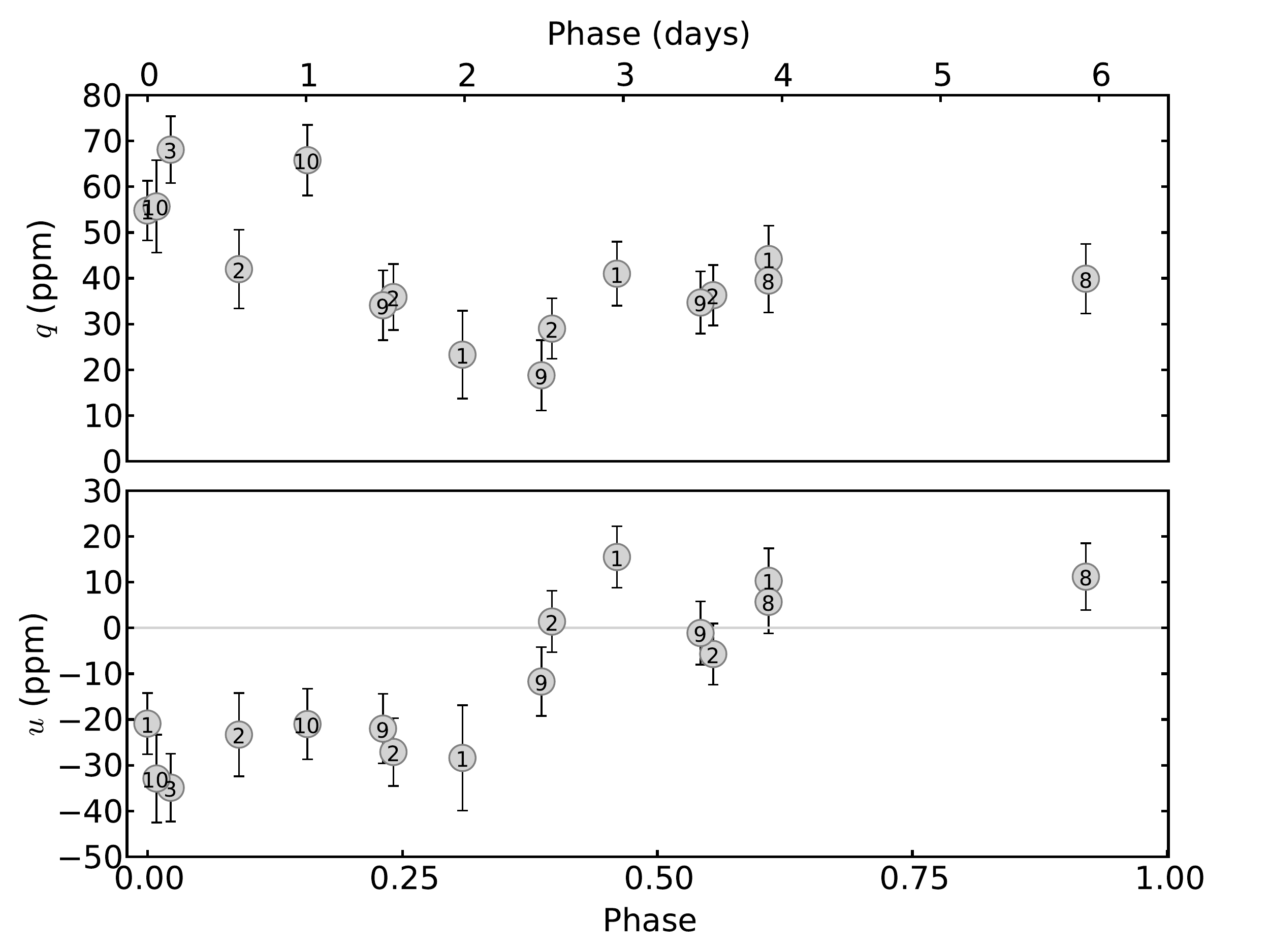}
\caption{Phase-folded behaviour of $\ksi$~Boo~A in $q$ (top) and $u$ (bottom), assuming a rotational period of 6.43~d. Numbers indicate the rotational cycle. It can be seen that there is excellent agreement in both $q$ and $u$ between cycles, even after 10 cycles.}
\label{fig:qu}
\end{figure*}

We phase-fold and plot the HIPPI data over the 6.43~d period in figure \ref{fig:qu}. We choose the first HIPPI observation as the epoch (UTC: 2017-06-22 11:34:57, JD: 2457926.9826). Data from parts of ten rotation cycles are shown, with points enumerated by cycle. The later and earlier cycles match very well, confirming rotational modulation. 

The shapes of the two fitted phase curves in figure \ref{fig:ml} are notably similar, implying a common origin. If the polarimetric mechanism is the global magnetic field of the star, then in the weak field case, the transverse magnetic field strength ($B_t$) will be proportional to $\sqrt{p}$ \citep{stenflo13}. By definition $B_t$ and $B_\ell$ are orthogonal, so that a point at the centre of the stellar disk contributing solely to $B_\ell$, will contribute solely to $B_t$ when at the limb. Consequently, $B_\ell$ and $B_t$ should be out of phase in stellar rotation by $\sim$0.25 -- as suggested by figure \ref{fig:ml}. Field fine structure complicates the picture; the relationship will not be precise because $B_\ell$ and $B_t$ are global properties, and at any given phase it is not the same stellar disk presented to the observer. 

\begin{figure*}
\centering
\includegraphics[width=0.75\textwidth,trim={0.75cm 0 -1.75cm 0}]{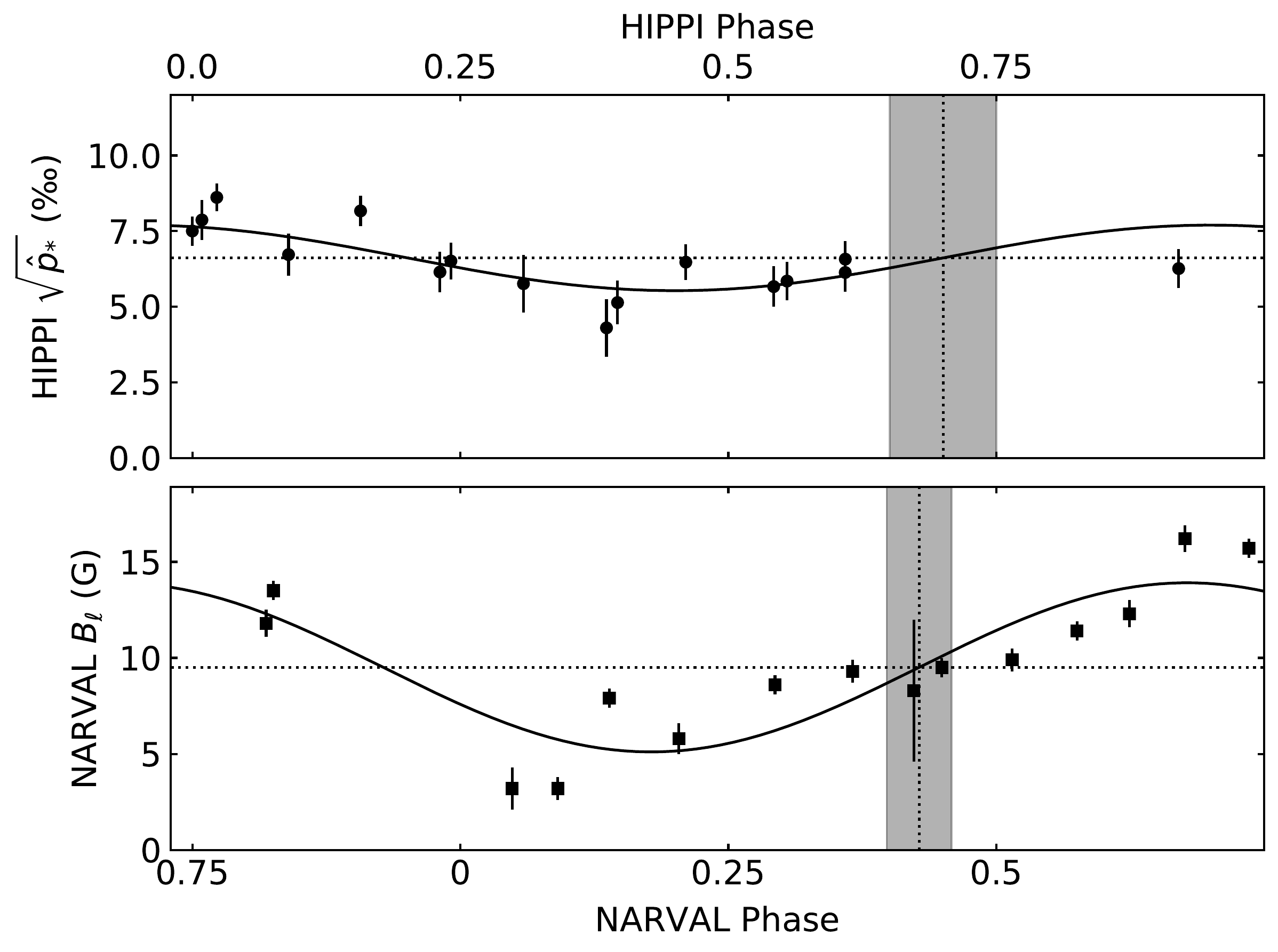}
\caption{Top panel: phase-folded intrinsic linear polarization measured in $\ksi$~Boo~A with HIPPI. Lower panel: the global longitudinal magnetic field ($B_\ell$) measured with NARVAL, offset in phase by 0.25. Shown on the plots are fitted sinusoids (black lines), fitted mean signals (dashed horizontal lines), fitted phases (dashed vertical lines) and their 1-$\sigma$ errors (grey panels). See also table \ref{tab:phase}.}
\label{fig:pv}
\end{figure*}

We fit a sinusoid with the equation: \begin{equation}\label{eq:sinusoids}y(t) = y_0 + A\sin 2\pi \left(t - t_0\right) / {T},\end{equation} of fixed period $T=$~6.43~d separately to each of the two datasets: $\sqrt{\hat{p_{\star}}}$ (assumed $\propto B_t$) and $B_\ell$, by allowing the offset $y_0$, the amplitude $A$ and the phase $t_0$ to vary. The fit coefficients and their uncertainties are shown in table \ref{tab:phase}. The phase of the HIPPI polarization measurements is 4.50~$\pm$~0.32~d which agrees well with the quarter-shifted NARVAL phase of 4.40~$\pm$~0.23~d as shown in figure \ref{fig:pv}. To provide a simple check of this finding we cross-correlated the maximum likelihood fit to the NARVAL data with the square root of the time-shifted first cycle of the HIPPI fit (i.e. those shown in figure \ref{fig:ml}), after first taking the minimum values as zero and normalising. This gave a maximum for a shift of $-$4.64~d, equivalent to a forward shift of $\sim$0.28 in phase units.

The ratio of $\sqrt{\hat{p_{\star}}}/B_\ell$ in $y_0$ is $\sim$0.68, which implies the efficiency of differential saturation. Naively one might expect that the ratio $\sqrt{\hat{p_{\star}}}/B_\ell$ would be the same for $y_0$ and $A$; figure \ref{fig:pv} shows clearly it is not. In fact, from table \ref{tab:phase} $A/y_0$ is $\sim$$\frac{1}{2}$ in the NARVAL data but only $\sim$$\frac{1}{6}$ in the HIPPI data. The relative amplitude in $\sqrt{\hat{p_{\star}}}$ is depressed compared to $B_\ell$. 

\begin{table}
\caption{Sinusoidal fitting of polarization measurements by equation \ref{eq:sinusoids}. The vertical offset, $y_0$, and amplitude, $A$, are $\sqrt{\hat{p_\star}}$ (\textperthousand) for HIPPI, and $B_\ell$ (G) for NARVAL. The phases are given both in days ($t_0$) and in phase units ($\varphi_0$).}
\tabcolsep 3.0 pt
\centering
\begin{tabular}{lcccc}
\hline
Dataset &  $y_0$ &    $A$ &  $t_0$ &  $\varphi_0$ \\
\hline
HIPPI  &   6.61~$\pm$~0.22 &   1.08~$\pm$~0.28 &   4.50~$\pm$~0.32 &         0.70~$\pm$~0.05 \\
NARVAL &   9.68~$\pm$~0.67 &   4.44~$\pm$~0.88 &   2.79~$\pm$~0.23 &         0.43~$\pm$~0.04 \\
\hline
\end{tabular}
\label{tab:phase}
\end{table}

\section{Discussion}
\label{sec:discussion}

The linear polarization measured in $\ksi$~Boo~A is rotationally modulated and 0.25 out of phase with the longitudinal magnetic field measured by spectropolarimetry, something which would not be expected from a strong field localised within starspots on the surface \citep{huovelin91}, nor hotspots \citep{schwarz84,clarke84}. So, the bulk of the linear polarization signal is due to the weaker global magnetic field. Differences in the fine detail of the two data sets may be easily explained by field fine structure, but might also be in part due to evolution of the field between the acquisition of the two data sets. We cannot rule out a small contribution from starspots breaking the disk symmetry. However, modelling by \citet{kostogryz15c} shows that for a star with a G7V spectral type, less than $\sim$1~ppm is expected from this process. Even for a cooler star with a spot covering 5\% of the disk, this will not exceed a few ppm. Only in very densely spotted stars will it be higher \citep{yakobchuk18}.

The suppression of linear polarization variations relative to what might be expected from $B_\ell \propto \sqrt{p}$ is probably a consequence of geometry. If the largest contributions to the net magnetic field are radial, then the strongest contributing surface regions to $B_t$ will be at the limb where the intensity is reduced. \citet{leroy90} describes this situation when considering relatively strong fields in starspots. \citep{huovelin91} made calculations that show that for a single small spot the linear polarization is greatest when the spot is 45 degrees from the disk centre (toward the limb); as larger spots are considered this approaches 90 degrees for spots up to 50\% of the surface area -- this being somewhat analogous to a global field. However, no specific modelling of differential saturation has been done for weaker global fields, and it is possible other magneto-optical phenomena may play a role. For instance, if the polarization scales non-linearly with field strength \citep{stift97}. Nevertheless, if this result is transferrable to other systems then broadband linear polarimetry is likely to be less sensitive to rotation than to large-scale changes in the field strength. 

With HIPPI we have a superior sensitivity to $q$ and $u$ Stokes parameters than spectropolarimetry \citep{cotton17a}, and we easily detect the time-varying magnetic field in $\ksi$~Boo~A. In the past broadband linear polarimetry has been used in conjunction with circular polarimetry to more precisely describe the magnetic fields of A- and B-type stars (e.g. \citealp{wade96}). Similar studies may now be made of the magnetic fields of cool stars. Alternatively, magnetic field maps produced with spectropolarimetry might be used to predict the associated broadband linear polarization for the purposes of subtracting it to reveal other phenomena. For this application either simultaneous measurements or a stable field would be required. More generally, the magnitude of linear polarization to expect in an active star might be predicted from a determination of $B_\ell$ if models are developed to describe differential saturation in a star with a weak global magnetic field. Follow up work is needed to develop such models. In particular the influence of spectral type should be examined in light of \citet{saar93}'s prediction of greater polarization in later spectral types where there are more spectral lines.

Searching for polarization from Rayleigh scattering particles in a hot Jupiter atmosphere in the combined light of the star and planet is a difficult prospect \citep{wiktorowicz15, bott18b}. In the best case, an orbital period-modulated signal will have an amplitude of up to a few 10s of ppm \citep{seager00, bailey18}. Such close systems are rare, of those known we estimate only 10--20 are bright enough to be accessible from large telescopes. Activity has the potential to obfuscate any planetary signal. As an example, consider $\tau$~Boo, which has an expected planetary signal of up to 8~ppm: in their linear polarimetric observations of the system \citet{lucas09} noted that the data displayed increased scatter, and suggested activity could be the cause. $\tau$~Boo has a value of |$B_\ell$|$_{max}$ of just 4.6~$\pm$~0.4~G \citep{fares09,mengel16}, but if the relationship found here for $\ksi$~Boo~A holds this could result in a mean linear polarization of 10-20~ppm, albeit with a periodic variability only some fraction of that.

Many other promising close hot Jupiter systems are known to be active, including HD~179949 \citep{fares12}, WASP~121 \citep{delrez16}, WASP~19 \citep{Huitson13}, and $\upsilon$~And and HD~209458 \citep{shkolnik05}. However, the system for which this result is most relevant is HD~189733, which has been favoured by groups looking for hot Jupiter polarization \citep{berdyugina11,wiktorowicz15b,bott16}. HST albedo measurements revealed the planet's blue colour \citep{evans13} making it a promising target, with an expected peak polarization of $\sim$20~ppm \citep{bailey18}. Yet it is known to be a fairly active BY~Dra variable with a field that evolves over time \citep{fares17}. Measurements made by \citet{petit14} give a |$B_\ell$|$_{max}$ value of 17.3~$\pm$~0.7 G (see also \citealp{fares17}), which could easily translate to a linear polarization of many tens of ppm. Indeed, \citet{bott16} describe a polarization offset for HD~189733 larger than expected for interstellar polarization. As that data comes from just a few short runs, stellar activity could easily be the cause. Clearly, it will be necessary to consider strategies for dealing with the star's activity if a definitive polarimetric detection of the planet is to be made. At the very least, data sets separated in time should not be combined without allowing for the potential of long term changes in the magnetic field producing an offset. 

\section{Conclusions}
\label{sec:conclusions}

We have observed rotationally modulated broadband linear polarization in $\ksi$~Boo~A with a characteristic variability of 10--14~ppm. The signal is out of phase by 0.25 with contemporaneous circular spectropolarimetric measurements of the longitudinal global magnetic field. We conclude that the broadband linear polarization is induced by differential saturation of spectral lines in the transverse global magnetic field, however the relationship is not as simple as $B_\ell \propto \sqrt{p}$, probably as a result of geometric considerations. Simultaneous measurements with spectropolarimetry could help constrain field geometry in cool stars, with further work to understand differential saturation in stars with weak global magnetic fields. 

The mean magnitude of the linear polarization is $\sim$4~ppm/G. If this is similar for other active stars, magnetism will complicate the search for polarization from close-in hot Jupiters in many exoplanet systems. Strategies will need to be employed to mitigate the effects; offsets might be applied to distinct data sets separately, or signals corresponding to the stellar rotation period sought and removed. Simultaneous observations with spectropolarimetry might allow the signal component due to the magnetic field to be removed directly, once the phenomenon is better understood. 

\section*{Acknowledgments}

This work was Australian Research Council supported through grant DP160103231. We thank the Director and staff of the Australian Astronomical Observatory, and acknowledge the BCool team's role in acquiring the spectropolarimetric data. This work has made use of the VALD database, operated at Uppsala University, the Institute of Astronomy RAS in Moscow, and the University of Vienna. We acknowledge the use of the SIMBAD database. AAV acknowledges funding received from the Irish Research Council Laureate Awards 2017/2018. SVJ acknowledges the support of the German Science Foundation (DFG) Research Unit FOR2544 ``Blue Planets around Red Stars'', project JE 701/3-1 and DFG priority program SPP 1992 ``Exploring the Diversity of Extrasolar Planets (RE 1664/18)''.

\small\bibliographystyle{mnras}
\bibliography{ksi_Boo}

\bsp

\appendix

\label{lastpage}

\end{document}